\documentclass[letterpaper]{article}
\usepackage{aaai}
\usepackage{subfig}
\usepackage{times}  
\usepackage{helvet}  
\usepackage{courier}
\usepackage{enumitem}
\usepackage{listings}
\usepackage{array}
\usepackage{mdwlist}
\usepackage{tabularx}  
\usepackage{tkz-2d}

\usepackage{tikz}

\usepackage{amsmath}
\usepackage{algorithm}  
\usepackage{algorithmic} 
\usepackage{multirow}
\usepackage{soul}
\newtheorem{theorem}{Theorem}

\newcommand{\lgraph}{\textsc{l$_{3}$}}
\newcommand{\kgraph}{\textsc{k$_3$}}

\title{A Hierarchical Dynamic Programming Algorithm for Optimal Coalition
Structure Generation}
\author{Meritxell Vinyals, Thomas Voice, Sarvapali Ramchurn, Nicholas R. Jennings\\
School of Electronics and Computer Science\\
 University of Southampton, UK\\
}
\begin{document}
%Efficient scalable supply chain formation: a message-passing approach
%Scalable high performance supply chain formation
%Scalable 
% TITLE DYNAMIC PROGRAMMING, PSEUDOTREES, SYNERGY GRAPHS, GRAPHS
%\title{An efficient scalable DP algorithm for Coalition Formation}
%\title{A Pseudotree-based representation for efficient Coalition Formation in
%Synergy Graphs}
%\title{Efficient Coalition Formation in Synergy Graphs based on Pseudotrees}
%\title{An efficient pseudotree-based representation of Graph-restricted
%Coalition Structure in Synergy graphs} 
%\title{An efficient dynammic programming representation for
%Graph-restricted Coalition Formation on Synergy Graphs}
%On efficient Coalition Formation Generation in Synergy Graphs
%PEDyPE
%PEDyCE
%EPDyCE.
%DyPCE
%EDyPCE
% \title{An efficient Dynamic Programming Algorithm for Coalition
%Structure Generation}
%\title{Pseudotree-based Coalition Structure Generation Algorithm}

\maketitle 
\begin{abstract}
\begin{quote} 
We present a new Dynamic Programming (DP) formulation of the
Coalition Structure Generation (CSG) problem based on imposing a hierarchical
organizational structure over the agents.
We show the efficiency of this formulation by deriving DyPE, a new
optimal DP algorithm which significantly outperforms current DP approaches in speed and memory usage.
% We show how this new formulation leads to a
%more efficient DP implementation, by proposing DyPE an optimal 
% In this paper we propose DyPE, an efficient optimal DP
%algorithm for Coalition Structure Generation (CSG). 
%DyPE operates on a novel
%representation of the CSG problem as a pseudotree. 
In the classic case, in which 
all coalitions are feasible, DyPE has half the memory
requirements of other DP approaches. On graph-restricted CSG, in which
feasibility is restricted by a (synergy) graph, DyPE has either
the same or lower computational complexity depending on the underlying graph
structure of the problem.
%Indeed,
%for trees, the memory and computational requirement of DyPE is linear in the
%number of agents and feasible coalitions respectively. 
Our empirical
evaluation shows that DyPE outperforms the state-of-the-art DP approaches
by several orders of magnitude in a large range of graph structures (e.g. for
certain scalefree graphs DyPE reduces the memory requirements by $10^6$
and solves problems that previously needed hours in minutes).
%For
%particular non-trivial classes of graphs, DyPE is the first algorithm to find
%the optimal coalition structure for problems with hundreds of agents in
% minutes.
\end{quote}
\end{abstract}

\vspace{-0.2in}\section{Introduction}

A key part of any coalition formation process involves partitioning the set of
agents into the most effective coalitions, (i.e. the optimal coalition
structure). However, this \emph{Coalition Structure Generation}
problem (CSG) is akin to the set
partitioning problem and hence NP-Hard~\cite{DBLP:journals/ai/SandholmLAST99}. 
%(hence NP-Hard), and scales in
%$\omega(n^n)$ \cite{DBLP:journals/ai/SandholmLAST99}. 
Over the last few years, several optimal CSG algorithms have been designed to
combat this complexity \cite{DBLP:journals/aamas/ServiceA11a,DBLP:conf/aaai/RahwanMJ12}.
In most cases, these algorithms were formulated for the classic CSG model
in which all coalitions are feasible. In contrast, in this paper, we tackle
the problem in which coalition membership is restricted by some kind of
(synergy) graph.
%many domains of interest the interaction of agents follows some structure.
%In particular, in this paper, we focus on problems where coalitional membership
%is restricted by some kind of (synergy) graph. 
Such restrictions have been widely studied in the context of cooperative
game theory \cite{greco_ijcai,demange2004} since they
naturally reflect many real-life settings, such as communication networks
\cite{myerson} and logistic networks \cite{johnson}.
% and group buying
%in social networks \cite{6348270}.
%Concretely, in this
%paper we assume that coalition membership is restricted by some kind of
% (synerg)

In these restricted settings, Dynamic Programming
(DP) approaches are attractive since they can solve the CSG problem by
simply assigning an infinite negative value to non-feasible coalitions.
%In the DP approach, the basic idea is to solve the
%CSG problem for a smaller subset of agents, using the answer of these
%subproblems to solve larger subproblems until solving the CSG among all
%agents. 
To date, all DP algorithms build on the same DP formulation of
the CSG problem, due to \cite{DPalgorithm}. 
As noted in \cite{DBLP:conf/atal/RahwanJ08}, this DP formulation leads to a
redundant search of the CSG space although some of this unnecessary calculations
are avoided by IDP, the fastest DP algorithm for classic CSG.
% IDP
%operates by avoiding unnecessary calculations to guarantee the optimal solution
% will be found.
% However, as already observed in \cite{DBLP:conf/atal/RahwanJ08}, this
%recursive formulation leads to a redundant search of the CSG space.
% This formulation is not
%efficient, in terms that it creates a lot of redundancies in the CSG search
% space.
%This recursive formulation is redundant.
%In DP there is an ordering among subproblems, and a
%relation that shows how to solve a subproblem given the answer of subproblems
% that appear earlier in the ordering.
%All current DP approaches build on the same recursive formulation of the CSG
% problem.
%This formulation is not efficient, in terms that it creates a lot of redudances
%in the search space.
%All current DP-CSG algorithms work by exploring
%subproblems by its size: subproblems corresponding to coalitions of size 1, to
%coalitions of size 2, and so on until reaching the grand coalition.
%In this class, the current state-of-the-art DP algorithm is IDP
% \cite{DBLP:conf/atal/RahwanJ08}, an improved DP implementation that avoids
% some parts of this redundant search. 
For (sparse) graph-restricted CSG, the
 fastest algorithm is DyCE \cite{DBLP:conf/aamas/VoiceRJ12}, that outperforms
 IDP by several orders of magnitude in sparse graphs by restricting the DP
 formulation to feasible coalitions in the graph.
 %Although IDP can deal with any feasible set of coalitions, the restriction on
%feasibility has no effect on its runtime.
%This, Voice et al. propose DyCE \cite{DBLP:conf/aamas/VoiceRJ12}, a DP
%approach specially designed for CSG problems where
%coalition potential membership is restricted by a (synergy) graph.  
%Commonly cited problems of this kind include communication network problems
%\cite{myerson}, logistic network problems \cite{johnson}, and, more recently,
% group buying in social networks \cite{6348270}.
Despite these advances, to date, the CSG problem can only be solved optimally
for up to 32 agents, even when considering graph restrictions
\cite{DBLP:conf/aamas/VoiceRJ12}.

%In this paper
%we focus on \emph{graph-restricted} CSG where coalition potential membership is
%restricted by a (synergy) graph.
% The idea of
%restricting coalitions to connected subgraphs of some graph is an old one, due
% to Myerson \cite{myerson} and widely

%Reduce
%To date, the best DP algorithm is IDP, an improved version of the original DP
%algorithm that reduces the time and  memory required when solving the CSG
%problem. More concretely, IDP achieves these savings with respect DP, by
%avoiding the exploration of some redundant paths that lead to the same
% coalition structure in the search space.

 Against this background, this paper presents DyPE, a new optimal DP algorithm
 which significantly outperforms current DP approaches and scales to larger
 problems.
DyPE operates on a novel formulation of the CSG problem that imposes a
hierarchical structure over the set of agents.
%In more detail, our
%contributions are presented as follows.
In more detail, this paper makes the following contributions:
%advances the
%state-of-the-art in the following ways:
 \vspace{-0.03in}\begin{itemize*}
  \vspace{0.015in}\item We introduce a new DP
  \emph{formulation} for the CSG problem that builds its search
  on a pseudotree hierarchy of the agents'
  synergy graph. We further show how this formulation enables a new search of
  the CSG space in which the coalitions an agent can join are conditioned on the
  coalitions formed by agents in earlier positions; \vspace{0.015in}\item We
  propose and prove the correctness of our new \emph{algorithm}, DyPE (\textbf{Dy}namic programming \textbf{P}seudotree-based optimal coalition structure \textbf{E}valuation) which is an efficient implementation of the hierarchical DP formulation;
  
  \vspace{0.015in}\item We analyse the \emph{complexity} of DyPE showing that
  it has either the same or lower computational
  complexity (depending on the structure of the synergy graph) than the current
  state-of-the-art DP algorithms;
  \vspace{0.015in}\item We \emph{empirically} show that DyPE solves the CSG
  problem faster than both IDP and DyCE in a range of graph structures, including the classic case
  (e.g. with a tree-restricted problem with 40 agents it is $10^4$ times faster
  and reduces by $10^7$ times the memory requirements). Moreover, for particular
  graph classes, DyPE solves the CSG problem for hundreds of agents in minutes.
\end{itemize*}
%First, we introduce a new recursive DP formulation for the CSG problem (with
% and without graph restrictions) that takes a hierarchy among agents to organize a
%novel exploration of the CSG search space. 
%Second, we propose and prove correct our new algorithm, DyPE (Dynamic
%programming Pseudotree-based optimal Coalition structure Evaluation) an
%efficient implementation of the recursive hierarchical formulation that uses as
%a hierarchy a pseudotree of the agents synergy graph. Third, we analyse the
%complexity of DyPE with respect to other DP algorithms for the CSG problem
%showing that DyPE has either the same or lower computational complexity
%depending on the underlying graph structure of the problem. Fourth, our
%empirical evaluation show that DyPE is able to solve the CSG problem, with and
%without graph restrictions, faster than both IDP and DyCE (e.g. for some
%scalefree graphs DyPE reduces $10^6$ the memory requirements of DyCE, solving
%problems for which DyCE needs hours in minutes). For particular non-trivial
%classes of graphs, DyPE is the first algorithm to solve the CSG problem for
%hundreds of agents in minutes.

\vspace{-0.02in}\noindent This paper is organised as follows. We proceed with a 
background section, followed by formulation, algorithm, complexity and empirical sections, and then conclusions.

\section{Background}
\label{sec:background}
\subsection{Basic Definitions}
\noindent Let
$A=\{1,\ldots,\vert A \vert \}$ be a set of agents.
A subset $C \subseteq A$ is termed a coalition. We denote the coalition composed
of all agents in $A$ as the \emph{grand coalition} and the coalition composed of
a single agent $i$ as its \emph{singleton}. A CSG problem is completely
defined by its characteristic function $v:
2^{A} \rightarrow \Re$ (with $v(\emptyset)=0$), which assigns a real value representing utility to every coalition. 
The CSG problem involves finding the exhaustive disjoint partition
of the set of agents into coalitions (or, \emph{Coalition Structure} (CS)) 
$CS=\{C_1,\ldots,C_k\}$ so that the total sum of values, $\sum^k_{i=1} v(C_i)$,
is maximised. 
%The CSG is known to be NP-Hard
%\cite{DBLP:journals/ai/SandholmLAST99}.

Now consider a CSG problem in which not all coalitions are assumed
feasible. Rather their feasibility is restricted by a \emph{synergy}
graph $G=(A,E)$ where:
(i) each node of the graph represents an agent; and (ii) a coalition $C$ is allowed to form
iff every two agents in $C$ are connected by some path in the subgraph induced
by $C$.\footnote{It is noteworthy that this representation subsumes
classic CSG: any classic CSG problem can be modeled as a graph-restricted
one by assuming a complete graph among agents.}
We denote the
set of feasible coalitions in a $G$-restricted CSG problem as $F(G)$.
The characteristic function of a $G$-restricted CSG problem returns minus
infinity for any non-feasible coalition $\forall C \not \in F(G): v(C)=-\infty $.
We will denote as \kgraph \ the complete graph among three agents  and  as
\lgraph \ the graph $G=(\{1,2,3\},\{\{1,2\},\{2,3\}\})$ in which the three agents interact in a line. Then, in a \lgraph -restricted CSG problem agent $1$ can not
form a coalition with agent 3 without agent $2$ (e.g. $C= \{1,3\} \not\in
F(G)$).
%Also we will use $T$ to
%denote an acyclic graph (a tree).

% \begin{example}[\lgraph -restricted CSG
%\label{ex:graph_example}] In a 3-line agent synergy graph
% $G=(\{1,2,3\},\{\{1,2\},\{2,3\}\})$ the tree agents interact in a line and agent $1$ can not form a coalition with 3
%without agent $2$ (e.g. $C= \{1,3\} \not\in F(G)$).
%\end{example}

In preparation for the further description of DP algorithms, we will
denote, for any subset $C\subseteq A$, $\Phi(C)$ as the
connected components of the induced subgraph of $C$ on $G$ and
$\Pi^C_k$ as the set of all partitions of $C$ into $k$ parts.

\subsection{Existing DP Approaches for CSG}

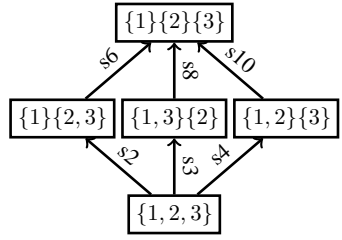
\begin{figure}[!tb]
 \hspace{-0.3in} \subfloat[ DP computations.\label{fig:operations}]{
 	\small
	\tikzstyle{node} = [rectangle,line width = 1pt, text
	centered,color=black,draw] \tikzstyle{arrow} = [line width = 1 pt]
	\begin{tikzpicture}[scale=0.8,transform shape]
	\node (a) at (-2.4,-2.75) {};
	\node (b) at (-2.4,2.75) {};
	\node (0) at (0,0) {
	\begin{tabular}{ l l p{0.07in}  }
	\hline
	$\mathbf{P}[C]$ &  $\mathbf{S_C}$  & $\#$ \\
	\hline
	$P[1]$ &  $v(1)$ & s13\\
	\hline
	$P[2]$ &  $v(2)$ & s12 \\
	\hline
	$P[3]$ &  $v(3)$ &s11\\
	\hline
	$P[1,2]$  &	 $v(1,2)$ &  s9\\
	 		  & $P[1] + P[2]$ & s10  \\
	\hline
	$P[1,3]$ 	&  $v(1,3)$ & s7\\
	 		  &	$P[1] + P[3]$ & s8 \\	
	\hline
	$P[2,3]$ 	&  $v(2,3)$ & s5 \\
	 			&  $P[2] + P[3]$  & s6 \\	
	\hline
	$P[1,2,3]$ 	& $v(1,2,3)$ & s1 \\
				& $P[1] + P[2,3]$ & s2 \\
				&  $P[1,3] + P[2]$ & s3 \\
	 			&  $P[1,2] + P[3]$ & s4  \\

	\hline
\end{tabular}
	};
	\path[<-]	(a) edge[arrow,sloped, above] node {Enumeration} (b);
	\end{tikzpicture}
	}
\subfloat[CS graph. \label{fig:csgraph}]{ \tikzstyle{node} =
[rectangle,line width = 1pt, text centered,color=black,draw] \tikzstyle{arrow} =
[line width = 1 pt] \begin{tikzpicture}[scale=0.85,transform shape] \node[node]
(0) at (0,0) {$\{1\} \{2\} \{3\}$}; \node[node] (1) at (0,-1.5) {$\{1,3\}\{2\}$};
	\node[node] (2) at (-1.75,-1.5) {$\{1\}\{2,3\}$};
	\node[node] (3) at (1.75,-1.5) {$\{1,2\}\{3\}$};
	\node[node] (4) at (0,-3) {$\{1,2,3\}$};
	\node at (0,-4) { };
	\node at (3,0) { };
	\node at (-3,0) { };
	\path[<-]   (0) edge[arrow,sloped, above] node {s8} (1)
				(0) edge[arrow,sloped, above] node {s6} (2)
				(0) edge[arrow,sloped, above] node {s10} (3)
				(1) edge[arrow,sloped, above] node {s3} (4)
				(2) edge[arrow,sloped, above] node {s2} (4)
				(3) edge[arrow,sloped, above] node {s4} (4);
	\end{tikzpicture}
	\label{fig:graphexampleb}
	}
	\label{fig:execution}
	\vspace{-0.1in}
	\caption{ a) Sequences of computations performed by DP and b) CS graph given
	$A=\{1,2,3\}$. }
	 \vspace{-0.2in}
\end{figure}

DP solves optimization problems recursively: 
a problem is solved by independently solving a collection of subproblems.
In CSG, a subproblem $P[C]$ stores the value of the best CS that can be
formed among agents in $C\subseteq A$.
%\begin{center}
% $P[C] = $ the value of the best CS that can be formed
%between agents in $C$.
%\end{center}
%\noindent The results of the
%subproblems are stored in a table, $P[\cdot]$, and its solution is reused on
%solving multiple subproblems. 

\noindent Current DP algorithms compute $P[C]$ by splitting the CSG search space
into $n$ subspaces ($\mathbf{S_C}$): one containing coalition $C$ and one for each
partition of $C$ into two sets $C_k$, $C_l$:
\begin{equation}
P[C] = \max \left( v(C),\max_{(C_k,C_l) \in \Pi^C_2 } P[C_k] + P[C_l] \right)
\label{eq:current_formulation}
\end{equation}

%Notice that the subspace corresponding to a
%partition $C_k$, $C_l$ contains all CS in which agents in $C_k$, $C_l$ are in
%different coalitions. 
Figure \ref{fig:operations} shows the trace of this DP formulation
specifying the set of evaluated subproblems ($P[C]$),
the subspaces evaluated for each subproblem ($\mathbf{S_C}$) and the number of
subspace ($\#$) over the classic CSG among three agents.

% Thus, whereas for the classic CSG
%among three agents DyCE will execute all operations detailed in Figure
%\ref{fig:operations}, for the CSG problem restricted by the 3-line graph from
%Example \ref{ex:graph_example}, DyCE discards the evaluation of subproblem
%$P[1,3]$ as well as of the subspace $s5$ since both involve
%$\{1,3\}\not\in F(G)$.
 
% The work in \cite{DBLP:conf/aamas/VoiceRJ12} builds on this observation to
%propose DyCE, a DP algorithm specially designed for graph-restricted CSG
%problems. More specifically, DyCE follows the recursive formulation given in
%Equation \ref{eq:current_formulation} but restricting it to coalitions that are
%feasible in the graph ($C_k ,C_j \in F(G)$). 
% In this case the CSG space can be divided into the coalition structure
%$\{ 1,2,3\}$ and three subspaces: s1 containing all CS in which $2,3$ does
%not join $1$;  s2 containing all
%CS in which 1,3 does not join $2$; and s3 containing all CS in which agents
%$1,2$ does not join $3$. The value of $P[\{1,2,3\}]$ is the maximum among the
%values of these four cases. Notice that finding the best CS in each subspace
%involves solving two subproblems. 
%Thus, for example, the solution of s1 involves computing the solution
% $P[\{1\}]$ and $P[\{2,3\}]$.

%In general, current DP algorithms solve a subproblem $P[C]$ using the following
%recursive formula:
To make sure that a
problem is computed before its subproblems in
Equation \ref{eq:current_formulation}, DP algorithms iteratively compute subproblems by size:
subproblems corresponding to all coalitions of size 1, to all coalitions of size
2, and so on until it reaches the grand coalition (note the direction of the
enumeration arrow in Figure \ref{fig:operations}). 

IDP algorithm by \cite{DBLP:conf/atal/RahwanJ08} is an improved implementation
of Equation \ref{eq:current_formulation} that prunes the evaluations of some
subspaces that are proven to be redundant during the search. DyCE algorithm by
\cite{DBLP:conf/aamas/VoiceRJ12}, specifically devised for graph-restricted problems, implements a variant of this DP
formulation that restricts coalitions in Equation \ref{eq:current_formulation}
to be feasible in the graph ($C_k ,C_j \in F(G)$).
  Thus, in Figure \ref{fig:operations}, if the CSG problem is restricted to the
  \lgraph \ graph, DyCE omits the evaluation of subproblem $P[1,3]$, as well
  as of subspace $s3$, since both involve $\{1,3\}\not\in F(G)$ whereas IDP goes
  through all subspaces independently of the graph.

 %Given this, to date, the fastest DP algorithm for solving the classic CSG
 %problem is IDP \cite{DBLP:conf/atal/RahwanJ08}, an implementation of this
 %DP formulation that prunes the evaluation of some subspaces that are redundant
 %in the search. 
 % For graph-restricted problems, the fastest algorithm
 %is DyCE that restricts coalitions in Equation \ref{eq:current_formulation} to
 % be feasible in the graph ($C_k ,C_j \in F(G)$).
 
The operation of these algorithms is typically
visualized on the \emph{coalition structure (CS) graph}.
In this graph, nodes stand for coalition structures and, following Equation
\ref{eq:current_formulation}, an edge connects two coalition structures iff one
of the coalition structures can be obtained from the other by splitting one
coalition into two. Figure \ref{fig:csgraph} depicts the CS graph among
three agents with edges numbered with the number of the corresponding
subspace that generated it.

\section{A Hierarchical Formulation for CSG}
\label{sec:formulation}
\noindent This section presents a novel hierarchical DP formulation of the CSG
problem based on a pseudotree of the agents' synergy graph. This
pseudotree structure allows us to define a more efficient search of the CSG
space in which the coalitions an agent can join are explored
conditioned on the coalitions formed by agents in earlier positions in the hierarchy.
We further show how this search can be visualised in a particular graph of
coalition structures that we refer to as a \emph{hierarchical coalition
structure} (HCS) graph.

% This section presents our novel DP formulation that
%guides its search based on a hierarchical structure over agents in terms of a
% pseudotree of its synergy graph. We in which the possible Thus, for each agent the formulation will explore the possible coalitions she can join conditionated to the decisions of agents in earlier
%positions.
 
 %We will show how this hierarchical structure guides the DP of the search space
 % in which agents

%\textcolor{red}{Can yu give some insight into why proposing a hierarchical
%structure is a good way to go?}.
% We further show how the
%operation of this hierarchical formulation can be visualized on a graph of
%coalition structures showing an exploration of the CSG search space
%that differs from current DP implementations.

%\noindent This section presents a novel representation of the CSG search
%space in terms of a pseudotree of its synergy graph. We call this
% representation the \emph{Pseudotree Coalition Structure Graph}. We further show how this
%representation can be used to guide a novel DP approach for
%solving the CSG problem itself.

\subsection{Synergy Graph Pseudotree}
  
\begin{figure}[!tb]
\subfloat[PT for \kgraph. \label{fig:pseudotreesa}]{
	\tikzstyle{node} = [circle,line width = 1pt,text width=1em, text
centered,color=black,draw] 
	\tikzstyle{root} = [circle,line width = 2pt,text width=1em, text
centered,color=black,draw]
\tikzstyle{arrow} = [line width = 1 pt]
\tikzstyle{arrow2} = [line width = 0.5 pt]
	\begin{tikzpicture}[scale=0.6,transform shape, rotate=90]
		\node[root, rotate=-90] (0) at (0,4.5) {\large$2$};
		\node[node, rotate=-90]  (1) at (0,3) {\large$1$};
		\node[node, rotate=-90] (2) at (0,1.5) {\large$3$};
		\node at (0,6) {};
		\node at (0,-1) {};
		%\node[node] (3) at (0,0) {\large$4$};
		\path[<-]	(0) edge[arrow,sloped, above] node {} (1)
					(1) edge[arrow,sloped, above] node {} (2);
		\path[-]	(0) edge[arrow2,bend right,black!40, dashed] node {} (2);
					%(2) edge[arrow,sloped, above] node {} (3);
	\end{tikzpicture}
}      
\subfloat[PT for \lgraph. \label{fig:pseudotreesb}]{ 
\tikzstyle{root} = [circle,line width = 2pt,text width=1em, text
centered,color=black,draw]
\tikzstyle{node} = [circle,line width = 1pt,text width=1em, text
centered,color=black,draw] \tikzstyle{arrow} = [line width = 1 pt]
\tikzstyle{arrow2} = [line width = 0.5 pt]
	\begin{tikzpicture}[scale=0.6,transform shape, rotate=90]
		\node[node, rotate=-90] (0) at (-1,2) {\large$1$};
		\node[root, rotate=-90]  (1) at (-0.5,4) {\large$2$};
		\node[node, rotate=-90] (2) at (0,2) {\large$3$};
		%\node at (0,2) { };
		%\node at (0,-2) { };
		\path[<-]	(1) edge[arrow,sloped, above] node {} (0)
					(1) edge[arrow,sloped, above] node {} (2);

	\end{tikzpicture}
}
\vspace{-0.1in}
\caption{Pseudotrees for synergy graphs \ \kgraph \ and \lgraph .
\label{fig:pseudotrees}}  
\vspace{-0.15in}
\end{figure}
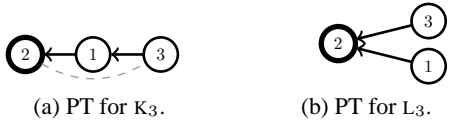

\noindent A pseudotree (PT) is a directed tree structure commonly used in search
and inference procedures \cite{DBLP:books/daglib/0016622}.
A pseudotree $PT$ of synergy graph $G$ is a rooted tree with agents $A$ as
nodes and the property that any two agents that share an edge in $G$ are on the
same branch in $PT$.
Here we restrict our attention to edge-traversal pseudotrees, namely those whose
edges correspond to edges in $G$. An edge-traversal pseudotree of a graph $G$ can be
computed by running a depth-first traversal search (DFS) algorithm
\cite{thesis_petcu}.
 Specifically, Figure \ref{fig:pseudotrees}
depicts two $PT$s where boldfaced edges are those included in the $PT$,
dashed edges are those in $G$ that are not included in the $PT$, and the
boldfaced node is the root agent.
Figure \ref{fig:pseudotreesa} shows a PT for the synergy graph of a
3-agent classic CSG problem rooted at agent $2$.
% Any PT of a complete graph takes the form of a line graph with root at one
% end.
Similarly, Figure \ref{fig:pseudotreesb} shows a pseudotree for the \lgraph \
graph rooted at agent $2$. Unlike in the \kgraph \ graph, agent $1$ and $3$ here
can be in different branches since they are not directly connected by an edge.

%A pseudotree categorises agents into levels, where the level of an agent is
%defined as the length of the path between $i$ and the root (the root agent has
%level $0$). 
%We also define the ancestors of an agent $i$ in the $PT$, $An(i)$,
%as all the agents in the path between $i$ and the root. 
We define the \emph{ancestors} of an agent $i$ in the $PT$
as all the agents in the path between $i$ and the root.
%Given a PT, let
%$L(\cdot)$ be a function that given an agent $i$ returns its level in the $PT$,
%namely its number of ancestors in $PT$ (the root agent has
%level $0$). In a CSG problem, a PT not only defines a partial ordering among
% agents but also on the set of feasible coalitions.
%Let $i$ be the agent with lowest level included in $C$. Notice that since $C$
% defines a connected subgraph in $G$ this agent is unique. Then we define the level of a coalition $C$ in
%$G$, namely $L(C)$, as the level of this agent in $PT$. %
%Formally, $L(C) = \min_{i\in C} L(i)$. 
Then, a pseudotree defines a partial hierarchical ordering among agents in which
any agent should be placed before any of its ancestors in the graph.
We will denote $O_{PT}$ as one ordering for pseudotree $PT$.
Notice that for the $PT$ in Figure \ref{fig:pseudotreesa}, $O_{PT}=\{2,1,3\}$ is
the unique ordering that satisfies the $PT$. In
contrast, for the $PT$ in Figure \ref{fig:pseudotreesb}, $O_{PT} = \{2,1,3\}$ and $O_{PT} =
\{2,3,1\}$ are both valid orderings.
 
Yet, $O_{PT}$ not only defines an ordering
among agents but also on the set of feasible coalitions. Let $i$ be the agent
with lowest order included in a coalition $C$.
Then, we define the order of $C$, $O(C)$, as the position of
$i$ in $O$. Formally, $O(C) = \min\{iÊ\vert O(i)\in C\}$. Thus, given $O_{PT}=
\{2,1,3\}$ the order of $\{1,2,3\}$ is 1 whereas of $\{1,3\}$ is 2.

%Given this, Equation defines an ordering among subproblems: each subproblem
%depends on a set of subproblems of higher order.
%Higher ordering. Agents in multiple branches can be in partial ordering because
%they will not appear in the same coalition without an agent of lower order, so
%the effect is the same.

%  Given a pseudotree, we define the level of a coalition $C$ in
%$G$, namely $L(C)$, as the  among the levels of its agents in $PT$. 
%Formally, $L(C) = \min_{i\in C} L(i)$. 
%Then, we define the set of coalitions ancestors of a coalition $C$, as all the
%coalitions that contain any of the ancestors of any agent included in $C$.
%Formally, $An(C) = \bigcup_{i\in C} An(i)$.
% Observe that in the same way that the pseudotree $PT$ defined a
%partial ordering among agents, it also defines a partial ordering among
% feasible coalitions, namely coalitions should be placed in the ordering before their ancestors
%coalitions.

\subsection{Hierarchical DP Formulation}
%\textcolor{red}{As we will show in the next section, we can use this ordering
% to guide a novel formulation of the CSG search space.}
%This hierarchical pseudotree structure imposes some structure on the
%CSG search space: for each agent it explores alternative coalitions she can
% join conditionated to the decision of other agents earlier in the hierarchy already taken. according to the structure of This hierarchical pseudotree structure can guide a new formulation of the CSG
%problem in which for each agent explores ex conditiononated to the possible
%decision of agents in earlier positions.
\begin{figure}[!tb]
\hspace{-0.15in}\subfloat[Hierarchical DP.]{
	\tikzstyle{node} = [rectangle,line width = 1pt, text
	centered,color=black,draw] \tikzstyle{arrow} = [line width = 1 pt]
	\begin{tikzpicture}[scale=0.75,transform shape]
	\node (a) at (-2.4,-2) {};
	\node (b) at (-2.4,2) {};
	\node (0) at (0,0) {
	\begin{tabular}{ l l p{0.05in}}
	\hline
	$\mathbf{P}[C]$ & $\mathbf{S_C}$ & s\# \\
	\hline
	$P[3]$ & $v(3)$ & s8\\
	\hline
	$P[1]$ & $v(1)$ & s7\\
	\hline
	$P[1,3]$ 	& $v(1,3)$ & s5\\
	 			& $v(1) + P[3]$ & s6\\		
	\hline
	$P[1,2,3]$ 	& $v(1,2,3)$ & s1 \\
	 			& $v(2) + P[1,3]$ & s2\\
	 			& $v(1,2) + P[3]$ & s3\\
	 			& $v(2,3) + P[1]$ & s4\\		
	\hline
\end{tabular}
	};
	\path[<-]	(a) edge[arrow,sloped, above] node {Enumeration} (b);
	\end{tikzpicture}
	\label{fig:dypecompletea}
	}
\subfloat[ HCS graph.]{
\tikzstyle{node} = [rectangle,line width = 1pt, text centered,color=black,draw] \tikzstyle{arrow} = [line width = 1 pt]
	\begin{tikzpicture}[scale=0.75,transform shape]
    %\node at (-2.85,0) {$L_2$};
	%\node at (-2.85,-1.5) {$L_1$};
	%\node at (-2.85,-3) {$L_0$};
	\node[node] (0) at (0,0) {$\{1\} \{2\} \underline{\{3\}}$};
	\node[node] (1) at (0,-1.5) {$\underline{\{1,3\}}\{2\}$};
	\node[node] (2) at (-1.75,-1.5) {$\underline{\{1\}}\{2,3\}$};
	\node[node] (3) at (1.75,-1.5) {$\{1,2\}\underline{\{3\}}$};
	\node[node] (4) at (0,-3) {$\underline{\{1,2,3\}}$};
	\path[<-]	(0) edge[arrow,sloped, above] node {s6} (1)
				(1) edge[arrow,sloped, above] node {s2} (4)
				(2) edge[arrow,sloped, above] node {s4} (4)
				(3) edge[arrow,sloped, above] node {s3} (4);
	\end{tikzpicture}
	\label{fig:dypecompleteb}
}
\vspace{-0.1in}
\caption{ a)  Hierarchical DP and b) HCS graph for the classic CSG with 
$O=\{2,1,3\}$.  Frontier coalitions are underlined.
\label{fig:dypecomplete}}
\vspace{-0.15in}
\end{figure}
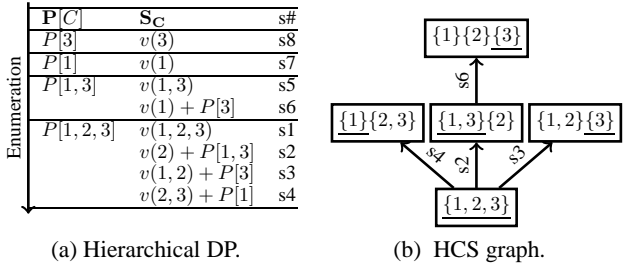

\begin{figure}[!tb]
\hspace{-0.15in}\subfloat[Hierarchical DP.]{
	\tikzstyle{node} = [rectangle,line width = 1pt, text
	centered,color=black,draw] \tikzstyle{arrow} = [line width = 1 pt]
	\begin{tikzpicture}[scale=0.75,transform shape]
	\node (a) at (-2.8,-1.5) {};
	\node (b) at (-2.8,1.5) {};
	\node (0) at (0,0) {
	\begin{tabular}{ l l p{0.05in}}
	\hline
	$\mathbf{P}[C]$ & $\mathbf{S_C}$ & s\# \\
	\hline
	$P[3]$ & $v(3)$ & s6\\
	\hline
	$P[1]$ & $v(1)$ & s5\\		
	\hline
	$P[1,2,3]$ 	& $v(1,2,3)$ & s1\\
	 			& $v(2) + P[1]+P[3]$ & s2\\
	 			& $v(1,2) + P[3]$ & s3\\
	 			& $v(2,3) + P[1]$ & s4\\		
	\hline
\end{tabular}
	};
	\path[<-]	(a) edge[arrow,sloped, above] node {Enumeration} (b);
	\end{tikzpicture}
	\label{fig:dypelinea}
	}
\subfloat[HCS graph. ]{
	\tikzstyle{node} = [rectangle,line width = 1pt, text
	centered,color=black,draw] \tikzstyle{arrow} = [line width = 1 pt]
	\begin{tikzpicture}[scale=0.75,transform shape]
	%\node at (-2.85,-1.5) {$L_1$};
	%\node at (-2.85,-3) {$L_0$};
	\node[node] (0) at (0,-1.5) {$\underline{\{1\}} \{2\}\underline{\{3\}}$}; 
	\node[node] (2) at (-1.8,-1.5) {$\underline{\{1\}}\{2,3\}$}; 
	\node[node] (3) at (1.8,-1.5) {$\{1,2\}\underline{\{3\}}$}; 
	\node[node] (4) at (0,-3) {$\underline{\{1,2,3\}}$}; 
	\path[<-]	(0) edge[arrow,sloped, above] node {s2} (4) 
				(2) edge[arrow,sloped, above] node {s4} (4)
				(3) edge[arrow,sloped, above] node {s3} (4);
	\end{tikzpicture}
	\label{fig:dypelineb}
	}
 \vspace{-0.1in}
\caption{ a) Hierarchical DP and b) HCS graph for the 3L-restricted CSG with 
$O=\{2,1,3\}$. Frontier coalitions are underlined. \label{fig:dypeline}}
\vspace{-0.15in}
\end{figure}
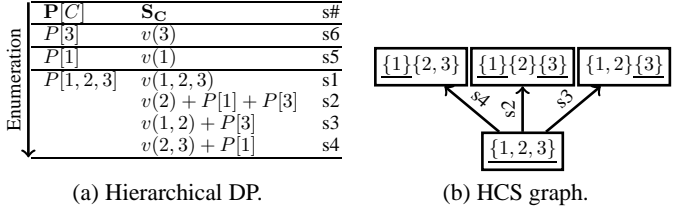

%Next we will illustrate by means of two simple examples how the
%the synergy pseudotree ordering introduced in section above can guide a new DP
%formulation to solve a CSG problem.
We next present the hierarchical DP formulation for classic CSG
illustrating its operation with two simple examples.

First, consider the classic CSG problem among three agents $A=\{1,2,3\}$.
Solving this problem involves comparing the value of five $CS$s (depicted as
nodes in the CS graph in Figure \ref{fig:csgraph}). Now, suppose that agents are
organised as in the PT of Figure \ref{fig:pseudotreesa} with an ordering
$O_{PT}=\{2,1,3\}$.
 %Our recursive formulation organises
%agents in the CSG organised into a hierarchy.
%Here, a hierarchy is a list of agents $O$ where $O(p)$ returns agent
%with order $p$ in the hierarchy.
%Thus, the hierarchy 2$>$1$>$3 is represented by $O= \{2,1,3\}$ where agent 2
%is the one with the lowest order. 
%In CSG, $O$ not only defines an ordering
% among agents but also on the set of coalitions. Let $i$ be the agent with lowest order included in $C$. 
%Then we define the order of a coalition $C$, namely $O(C)$, as the position of
%$i$ in $O$. Formally, $O(C) = \min\{iÊ\vert O(i)\in C\}$. Thus, given $O=
% \{2,1,3\}$ the order of coalition $\{1,2,3\}$ is 1 whereas of $\{1,3\}$ is 2.
%We will use this example to illustrate how this hierarchical ordering can
%be used to design a novel DP formulation for the CSG problem, with and without
%graph restrictions. 
Given the lowest agent in the ordering, agent $2$,
the CSG space can be divided into four subspaces, one for each \emph{feasible} coalition that
contains this agent; namely:
$s1$ containing $\{\{1,2,3\}\}$; $s2$ containing $CS$s that include $\{2\}$
and any CS among $1,3$; $s3$ containing $CS$s that include $\{1,2\}$ and any CS
among $3$; and $s4$ containing $CS$s that include $\{2,3\}$ and any CS among
$1$. Notice that finding the best $CS$ in each subspace involves solving a CSG
subproblem and that the agents in this subproblem depend on the particular
coalition that agent $2$ formed in this subspace.
% Notice that finding the best CS in each subspace involves computing
%the value of the CSG subproblem among the remaining agents not taken in the
% fixed coalition.
Thus, for example, the solution of $s2$ involves computing the best CS among
agents not present in coalition $\{2\}$, i.e. computing subproblem $P[1,3]$.
Again, we can solve this problem
by taking the agent with lowest order (agent 1), and repeating the
above process. Figure \ref{fig:dypecompletea} shows a complete trace of this
example. It is noteworthy that we reduced the
number of operations with respect to the current DP
operation: we solved 4 subproblems by evaluating 8 subspaces whereas
the current DP operation solves 7 problems by evaluating 13 subspaces (compare
Fig. \ref{fig:dypecompletea} with Fig. \ref{fig:operations}).
% on the 3-agent classic CSG problem with ordering $O=\{1,2,3\}$.
% that is agent $2$, and dividing the CSG space into two subspaces:
%one containing coalition $\{2,3\}$ and another containing CS's composed of
% coalition $\{2\}$ and any CS among $\{3\}$. Finally, finding the best CS among $\{3\}$ is trivial since it contains only its
%singleton. 
%\noindent As example consider the classic CSG problem among three agents, in
%which agents are placed in an hierarchical ordering $O=\{1,2,3\}$.
 %Solving this problem involves comparing the value of five CS's (depicted as
 %nodes the CS graph in Figure \ref{fig:graphexamplea}). 
 %Now, suppose
 %that agents are organised as in the $PT$ of Figure \ref{fig:pseudotreesa}
 %and take the agent with lowest level, namely $1$.
%Let $i$ bet the agent in $A$ with lowest level in the $PT$.
%In the PT of Figure, the
%agent with lowest level is $1$. 

This hierarchical DP search can be similarly applied to any
graph-restricted problem. For example, we can follow the same approach in the
CSG problem when restricted by the \lgraph \ graph (a complete trace is given in Figure \ref{fig:dypelinea}).
In this case when
computing subspace $s2$,  $P[1,3]$ can be decomposed into two independent
subproblems, namely $P[1]$ and $P[3]$, since agents $1,3$ do not
interact (are disconnected) given agent 2 formed a coalition without including
them. Again, note that this search is more efficient than the current DP
operation: it solves 3 subproblems by evaluating 6 subspaces where the
graph-restricted current DP operation 
solves 6 problems by evaluating 10 subspaces (compare Fig. \ref{fig:dypelinea}
with Fig. \ref{fig:operations}).
%when
%excluding agent 2.

%$P[1,3]$ can be computed as
%the sum of two subproblems, namely $P[1]$ and $P[3]$ since agents $1,3$ are
%disconnected when excluding agent 2. 
%In general, in graph-restricted CSG, the
%value of $P[C]$ is equal to the sum of the value of the
%subproblems corresponding to its connected components.
 
%Second, in graph-restricted CSG if we have two subsets of agents $C$ and $C'$
%which are disconnected then $P[C\cup C'] = P[C] + P[C']$. So, for any subset of
%agents $A$, the value of the subproblem $P[C]$ is equal to the sum of the value
%of the subproblems corresponding to its connected components. 
%As a
%consequence, the evaluation of subproblem $P[1,3]$ can also be omitted since
%it is not longer needed during the recursion.
%For example, if the interactions of agents in the CSG scenario
%describe above are restricted by the 3-line graph of Example
%\ref{ex:graph_example}, subspace $s3$ does not need to be evaluated. 

In general, given an ordering $O$ among agents, the solution of a $G$-restricted
CSG problem can be computed as comparing (maximising) over $n$ subspaces where
the value of each subspace can be solved by evaluating a \emph{feasible}
coalition and a set of subproblems corresponding to the connected
components\footnote{The connected components of a graph G are the set of
the largest subgraphs of G that are connected.} of the rest of agents not
present in this coalition.
%where $C$ has been restricted to $F(G)$ and the evaluation of subproblem
% $A\setminus C$ splits into the sum of its connected components.
Formally, we can define our hierarchical DP formulation for the
graph-restricted CSG problem as:
\begin{equation}
P[C]= \max_{\substack{C_k \subseteq C:\\ O(C_k) =O(C),\\  C_k\in F(G)}} \left(
v(C_k) + \sum_{C_l\in \Phi(C\setminus C_k)}P[C_l] \right)
\label{eq:new_formulation_final}
\end{equation}
%\noindent where $CC(S)$ returns the set of connected components of
%the induced subgraph of $S$ on $G$. 
%Notation: Here, for any set of agents $S$, we define
%$CC(S)$ to be the set of connected componentsof the induced subgraph of
%$S$ on $G$\textcolor{red}{example}.
%\begin{equation}
%P[A]=  \max_{\substack{C \subseteq A:\\  O(C) = O(A)}} \left( v(C) +
% P[A\setminus C] \right)
%\label{eq:new_formulation}
%\end{equation}
%Equation \ref{eq:new_formulation_final} defines a new DP formulation for
%the graph-restricted CSG problem, that we refer to as hierarchical DP
%formulation.
%Notice that this new formulation derives to different executions given
% different orderings among agents.

%In our case of interest, the feasibility of the coalitions are restricted by
% the graph, so $C$ in Equation \ref{eq:new_formulation} should be restricted to
%$F(G)$. 

%Accordingly, to represent explicitly graph-restriction feasibility Equation
% \ref{eq:new_formulation} changes to:

%\begin{equation}
%P[A]= \max_{\substack{C \subseteq A:\\ O(C) =O(A),\\  C\in F(G)}} \left( v(C) +
%\sum_{A'\in CC(A\setminus C)}P[A'] \right)
%\label{eq:new_formulation_final}
%\end{equation}
%where $C$ has been restricted to $F(G)$ and the evaluation of subproblem
%$A\setminus C$ splits into the sum of its connected components. Now, all
%evaluated coalitions and subproblems search by the recursion are feasible in
% the graph.

\subsection{Hierarchical Coalition Structure Graph}

%Similarly to current DP approaches, the operation of the hierarchical recursive
%formulation can be visualised using a graph,
%which we refer to as \emph{hierarchical coalition structure} (HCS) graph.
To discuss how to construct a HCS graph we need first to define the notion of
frontier coalitions.
Let us define the frontier coalitions of a coalition structure $CS$ as the set
of coalitions that are not connected to any other coalitions of higher order in $CS$. That is,
the frontier coalitions are the $C \in CS$ such that if $O(C') > O(C)$ for $C'
\in CS$ it implies $C \cup C'$ is disconnected.
%\begin{definition}
%\emph{Frontier} coalitions of $CS$ are these coalitions that are not connected
%to any other coalitions of higher order in $CS$: $C\in CS$ such that $O(C') >
%O(C)$ for $C'\in CS$ implies $C\cup C'$ is disconnected. 
%\end{definition}
In the HCS graph, nodes stand for feasible coalition structures and, 
following
Equation \ref{eq:new_formulation_final}, an edge connects two
feasible coalition structures if and only if one of the coalition structures can be obtained from the other 
by the evaluation of some subspace of one of its \emph{frontier coalitions}.
That is $CS$ is linked to $CS'$ if and only if there exists $C
\in CS$, $C' \in CS'$ with $C' \subseteq C$, $CS' = (CS \setminus \{ C \}) \cup
\{C', \Phi(C \setminus C')\}$ and $C'$ is a frontier coalition of $CS$.

% The
%set of \emph{frontier} coalitions of a coalition structure $CS$ are these
% coalitions that are not connected to any other coalitions of higher order in $CS$.

%As we will formally prove
%later, this non-redundancy is guaranteed in classic and $T$-restricted CSG. 

Figure \ref{fig:dypecompleteb} depicts the HCS graph of the 3-agent classic CSG
with $O=\{2,1,3\}$ and Figure \ref{fig:dypelineb} the graph when the
problem is restricted by the \lgraph \ graph.
Frontier coalitions are underlined in the graph. Notice that frontier coalitions
correspond to subproblems that would be evaluated during the operation of the hierarchical DP formulation.

\section{DyPE}
\label{sec:dype}

%\noindent In this section, we introduce the DyPE algorithm. First, we introduce
%the interaction pseudotree ordering, a hierarchy that builds on a
%pseudotree of the interaction graph.
%Then, we present the formal foundations for DyPE, namely how its implements the
%% hierarchical recursive formulation based on the interaction pseudotree
% ordering, and introduce its algorithmic details.
%Finally, we prove its correctness and
%its non-redundancy on particular graph classes.

\noindent We can now describe the operation of DyPE. First, we lay its
formal foundations, namely how its search is efficiently
derived from the hierarchical DP formulation, then we move to its algorithmic
details and, finally, prove its correctness.
%Then, we describe its algorithmic details and finally prove its correctness.
% a
%DP implementation of the hierarchical formulation for CSG that uses a hierarchy
%% based on a pseudotree of the synergy graph.
%Following sections formally describe
%DyPE and its algorithmic properties.

\subsection{Formal Foundations}
\label{sec:formalfoundations}

DyPE implements the hierarchical recursive
formulation introduced in the former section. Accordingly, DyPE:
(i) enumerates all subproblems in a bottom-up order (from subproblems that appear
  last in the recursion to the grand coalition); and (ii)
computes the value of each enumerated subproblem.

In the hierarchical
recursion formulation a subproblem is computed using the results of a set of subproblems
of higher order (a subspace of a subproblem $C$ contains a coalition of the same
order as $C$ and a set of subproblems of higher order).
Thus, to guarantee a valid exploration order (so no problem is evaluated before one of its subproblems) 
DyPE evaluates subproblems corresponding to feasible coalitions by its order,
from highest to lowest.
However, not all feasible coalitions are required during the recursion. 
For example, the DP execution in Figure \ref{fig:dypecompletea} requires
enumerating subproblems $\{3\}$, $\{1\},\{1,3\}$, and $\{1,2,3\}$ but
not $\{2,3\}$, although it
corresponds to a feasible coalition.

Detecting which of the subproblems will actually be
needed\footnote{Although a simple solution for only solving subproblems that are
actually needed is memoization \cite{dpbook}, the exponential overhead that
incurs recursion in this case makes it not applicable.} is
crucial for the performance of the DP implementation.
%(otherwise it will require to solve and store one subproblem ) to avoid solving
%and storing one subproblem for feasible coalition, which may be a very large
% number (even in sparse graphs).
DyPE exploits the fact that the ordering is based on a synergy pseudotree to
detect a necessary condition that any feasible coalition needs to satisfy in order to be
evaluated as a subproblem during the recursion. 
%In particular,  given a feasible
%coalition $C$, with lowest level agent $i \in C$, DyPE will evaluate $C$ as a
% subproblem if and only if all agents not in $C$ that lie on the branch of $PT$ rooted at $i$ are
%connected to an agent lower than $i$ through a path that does not pass through
%$C$. If $i$ is the root of $PT$, then this condition is only true for
%$C = A$. For other
%$i$, this condition is equivalent to saying that $A\setminus C$ is
%connected. The correctness of these conditions is formally proved in next
%sections, by proving the correctness of DyPE. 
In particular, let $C$ be a feasible coalition. If $C$ contains the root of
$PT$, DyPE will only evaluate $C$ if it is the grand coalition ($C=A$).
Otherwise, if $C$ does not include the root,
DyPE will only evaluate $C$ if the set of remaining agents, $A\setminus C$, is
connected. The correctness of these claims is formally proved in next
sections, by proving the correctness of DyPE.

\vspace{0.1in}\subsection{The Algorithm}
\label{sec:dypealgorithm}
\setlength{\textfloatsep}{0pt}% Remove \textfloatsep
\begin{algorithm}[!tb]
\caption{\textbf{\small DyPE( $v(\cdot)$, $A$, $G$, $O$)}} 
\small
\begin{algorithmic}[1]
%\FORALL{$i=1 \ldots \vert A \vert$ }
\STATE $C \leftarrow \emptyset;S\leftarrow \emptyset;$ /*Current subproblem,
current subspace*/ 
\STATE $i \leftarrow \vert A \vert;$ /*Start exploring the last agent
in the ordering $O$*/
\WHILE{$(C, i) \leftarrow$\textbf{nextSubproblem($C$, $i$)}}
%\FORALL{$C\subseteq A$ such that $\vert C \vert = s$ and $C$ is feasible /*For
%each feasible coalition of size s*/} 
	\STATE $P[C] \leftarrow -\infty;$ 
	\WHILE{$C' \leftarrow $  \textbf{nextConnectedSet}$(C', i, C)$  } 
	\STATE $V \leftarrow v(C') + \sum_{C''\in \Phi(C\setminus C')} P[C''];$
%\FORALL{$$}
%		\STATE $V \leftarrow V + P[S']; $ /*Add the value of subproblem
%		$S'$*/
%\ENDFOR
		\IF{$P[C] < V$ /*Compare the value of subspaces*/} 
			\STATE $P[C] \leftarrow V;$ /*Update subproblem value*/
			\STATE $B[C] \leftarrow C';$ /*Update the best subspace*/
		\ENDIF
	\ENDWHILE
\ENDWHILE
%\ENDFOR
\RETURN $bestCS(A);$
\end{algorithmic}
\label{alg:genericDP}
\end{algorithm}
\setlength{\textfloatsep}{5pt}% Remove \textfloatsep
%\floatname{algorithm}{Procedure}
\begin{algorithm}[!tb]
\caption{\textbf{\small nextSubproblem($C$, $i$)}} 
\small
\begin{algorithmic}[1]
\small
\IF{$i=1 $ /*For the root agent */}
	\IF{$C=\emptyset$}
		\RETURN ($A$, $i$); /*Only the grand coalition */
	\ENDIF
	\RETURN $\emptyset$; /*All agents explored, return empty set.*/
\ELSE
%\ENDIF
%\IF{$p < \vert A \vert $ /*For agent in position p  */}
	\WHILE{$C\leftarrow \textbf{nextConnectedSet}(C, i, \{i,
	\ldots, \vert A \vert \})$ }
	\IF{$A\setminus C$ is connected}
		\RETURN $(C, i)$;
	\ENDIF
	\ENDWHILE
	\RETURN \textbf{nextSubproblem}($\emptyset$,$i-1$); /*Recursively call with the
	previous agent in the ordering*/
\ENDIF
\end{algorithmic}
\label{alg:nextSubproblem}
\end{algorithm}

\noindent For notational convenience, 
we use \emph{nextConnectedSet$(\cdot, \cdot, \cdot)$} as an iterator function of
a connected subgraph enumeration (CSE) algorithm
\cite{DBLP:conf/aamas/VoiceRJ12,Gutin08analgorithm}.
That is, for any feasible coalitions $C' \subseteq C$ with $i \in C'$ \break
\emph{nextConnectedSet$(C', \{i\}, C)$}, returns the subset of $C$ that would follow $C'$ during the process of the chosen CSE algorithm as it iterates through all 
feasible subcoalitions of $C$ that contain $i$. If $C'$ is the last subset to be enumerated by the
CSE, the function returns the empty set.

%\textcolor{red}{To reduce the complexity of the analysis that follows in 
%Section \ref{sec:complexity}, we describe DyPE using several sub-algorithms.}
%Algorithm \ref{alg:genericDP} describes the main operation of the DyPE
% algorithm.
%In doing so it gives a general high level view of 
%the operation of the DP approach to solving a CSG problem.
%In addition, Algorithms \ref{alg:nextSubproblem}, and
%\ref{alg:evaluateSplitting}, called during the execution of Algorithm
%\ref{alg:genericDP}, outline the specific searches that DyPE executes
%during it's exploration of the PCS
%space (which distinguish it from other existing DP-CSG algorithms).
The pseudocode of DyPE is provided in Algorithm \ref{alg:genericDP}.
As can be seen, DyPE takes as an input a $G$-restricted CSG problem and an
ordering $O$ that satisfies a synergy pseudotree of $G$.
Let us assume, without loss of generality, that agents are numbered according to
an ordering that satisfies $PT$, so the root is $1$. 
After initialisation, DyPE proceeds to enumerate subproblems, 
using the iterator function \emph{nextSubproblem($\cdot, \cdot$)}.
For each agent $i$ in the ordering $O$, DyPE goes
through all subproblems that need to be evaluated where $i$ is the agent with
lowest order.

%Once this last subproblem has
%been evaluated, the iteration ends (line 5).

For each subproblem $C$, 
Algorithm \ref{alg:genericDP} computes the value of the
CSG problem over agents in $C$ (lines 4-11). 
To do so, it goes over all
subspaces of $C$ that need to be
evaluated ($\mathbf{S_C}$) by iteratively calling function
\emph{nextConnectedSet$(\cdot, \cdot,\cdot)$} (line 5).
In this way, DyPE evaluates one subspace
for each feasible subcoalition $C'$ of $C$ that contains agent $i$.
The value of the subspace is computed as the value of coalition
$C'$, $v(C')$, plus the value of each subproblem corresponding to each connected
component in $C\setminus C'$, $\Phi(C\setminus C')$ (line 6). The value of
$P[C]$ is computed in Algorithm \ref{alg:genericDP} as the maximum between the
values of the evaluated subspaces (lines 7-10).

%For each splitting $S$, the algorithm computes
%its value, $V_{S,C}$ by calling function \emph{evaluateSplitting(S,C)},
% updating the value of the subproblem iff applies (line ). 
%To summarize, from Algorithm \ref{alg:genericDP} we obtain that the value of a
%subproblem $C$ is computed as the maximum between the value of coalition $C$
% and the value of the set of evaluated splittings of $C$: $P[C] = \max(v(C),
%\max_{S\in \mathbf{\mathbf{S_C}}} V_S)$. 

At the end of this process, the
solution of the subproblem corresponding to the grand coalition $(P[A])$,
contains the value of the best CS explored during the execution of the
algorithm.
To recover the best CS, Algorithm \ref{alg:genericDP} also stores the subspace
that maximizes each subproblem $C$ in $B[C]$ (line 12) and at the end of its
execution calls a recursive procedure $bestCS(\cdot)$ over the grand coalition,
where $bestCS(C)$ returns $\{C\}$ if $C\setminus B[C] = \emptyset$;
$bestCS(B[C])\cup \bigcup_{C'\in \Phi(C\setminus B[C])}bestCS(C')$ otherwise.

The definition of \emph{nextSubproblem($\cdot, \cdot$)} is
given in Algorithm \ref{alg:nextSubproblem}. 
For agents $i=\vert A \vert \ldots 2 $ (i.e. excluding the root)
DyPE, uses \emph{nextConnectedSet$(\cdot, \cdot,\cdot)$} to enumerate
as subproblems every feasible coalition $C$ consisting of agent $i$ and any
subset of agents placed after $i$ in the ordering, $\{i, \ldots,
\vert A \vert \}$ (line 7).
DyPE evaluates subproblem $C$ only if the rest of the graph, $A\setminus C$,
remains feasible (lines 8-10). Lastly, for the root agent ($i=1$) , DyPE evaluates a single subproblem
corresponding to the grand coalition (lines 1-5). 
%outlined in Algorithm \ref{alg:valuePropagation}.
%For a relatively small computational overhead, the memory requirements of 
%Algorithm \ref{alg:genericDP} can be reduced by not
%storing the optimal splittings in $B[C]$, but instead finding them by searching
% through values of $P[\cdot]$, as in \cite{DBLP:conf/atal/RahwanJ08}.

%\begin{algorithm}[!tb]
%\caption{\textbf{\small bestCS($C$)}} 
%\small
%\begin{algorithmic}[1]
%\IF{$C\setminus B[C] == \emptyset$}
%	\RETURN $\{C\}$;
%\ELSE
	%\RETURN $bestCS(B[C]) \cup  bestCS(C\setminus
	%B[C])$;
	%\RETURN $bestCS(B[C])\cup \bigcup_{C'\in \Phi(C\setminus B[C])}bestCS(C');$
%\ENDIF
%\end{algorithmic}
%\label{alg:valuePropagation}
%\end{algorithm}

%Figure \ref{fig:dypecompleteb} show a trace of DyPE, one considering a
%general CSG among three agents. Figure \ref{fig:dypelineb} consider DyPE
%operation for the same CSG but restricted restricted by the 3-line
%synergy graph in Figure \ref{fig:graphexamplea}.

\vspace{-0.05in}\subsection{Correctness of DyPE}
\label{sec:properties}
\noindent The next theorem proves the correctness of DyPE.
%We now elaborate on two important properties of DyPE: 
%correctness (Theorem \ref{th:correctness}) and non-redundancy (Theorem
%\ref{th:efficiency}).
%First, note that DyPE only
%explores subproblems and splittings corresponding to feasible coalitions.
%So by splitting, we will never generate, nor explore, any coalition structure
%that contains an infeasible coalition.
% proves the \emph{correctness} of DyPE
%First, Theorem \ref{th:correctness} proves the \emph{correctness} of DyPE
%by showing that for any given graph-restricted CSG problem, DyPE
%returns an optimal coalition structure.
%Next, Theorem \ref{th:efficiency} shows how for complete and tree synergy
% graphs DyPE is guaranteed to non-evaluate any \emph{redundant} subspace (if we omit any
%of the evaluated subspaces then at least one of the feasible CS is not
% searched).
%This is equivalent to proving that DyPE fully explores the PCS graph for that
% CSG.
%Next, we will explore the \textcolor{red}{efficiency} of DyPE (Theorem
%\ref{th:efficiency}), and show that for complete and tree synergy
%graphs, all evaluated splittings are non-redundant (if we omit any of 
%the evaluated splittings then at least one of the feasible CS
%is not searched).

\vspace{-0.05in}\begin{theorem}[Correctness]
For any given graph-restricted CSG, DyPE calculates an optimal
coalition structure.
\label{th:correctness}
\end{theorem}
\textbf{Proof.} 
Since the value of the best coalition structure $CS^*$ returned by DyPE is equal
to $P[A]$, it is sufficient to show that, for every feasible coalition structure $CS$,
on completion of the algorithm, $P[A] \geq v(CS)$.

Given such $CS$, let $L$ be the order of the coalition with highest order
in $CS$. Let $\mathbf{C}_{\leq l}$ be the set of coalitions in $CS$ with order
equal or lower than $l$ ($\mathbf{C}_{\leq l} = \{C\in CS \vert O(C) \leq l\}$).
We prove the result by showing that for all $l=1 \ldots L$ 
$P[A] \geq V_l$ where $V_{l}$ contains the accumulated value (until step $l$) of
an exploration ``path'', \vspace{-0.05in}$$
V_l = \sum_{C \in \mathbf{C}_{\leq l} } v(C) + \sum_{C' \in \Phi(A \setminus
\mathbf{C}_{\leq l})} P[C'], $$\vspace{-0.1in}

\noindent and so $V_{L} = v(CS)$. We prove this by induction on $l$.

In the base case, $l=1$ and $\mathbf{C}_{\leq 1}$ is composed of a single
coalition, $C_1$, the coalition that contains the root agent in $CS$.
 Notice that all the subspaces of the grand coalition $A$ corresponding to
 feasible coalitions that contain the root are evaluated. Thus, the
 subspace $C_{1}$ of $A$ ($v(C_{1}) + \sum_{ C\in \Phi(A\setminus C_{1})} P[C]$)
 is evaluated and so $P[A] \geq V_1$.

In the inductive step, consider the coalition in $CS$ whose level is $l+1$,
$C_{l+1}$, (if there is a coalition with order $l+1$, this coalition is unique
since it is the one that contains the agent with order $l+1$)
%all coalitions in $CS$ whose level is $l+1$ ($\mathbf{C}_{=
%l+1}$) 
and that the induction hypothesis holds for all coalitions in $CS$ whose level is less or equal than $l$. Thus, $CS_{\leq l} = \{\mathbf{C}_{\leq l} ,
\Phi(A \setminus \mathbf{C}_{\leq l})\}$ is connected through a path to the grand
coalition $A$. Then, there must be one $CC \in \Phi(A \setminus
\mathbf{C}_{\leq l})$ such that $C_{l+1}\subseteq CC$. 
%For each coalition $C'\in
%\mathbf{C}_{= l+1}$ there must be some $CC' \in CC(A \setminus \mathbf{C}_{\leq
% l})$ such that $C'\subseteq CC'$. 
Since the ordering follows a pseudotree, the union of coalitions in
$\mathbf{C}_{\leq l}$ forms a connected subgraph,
$A\setminus CC$ must be feasible and thus, $CC$ is evaluated as a
subproblem.
As all agents in $CC$
have higher order than $l$, $O(C_{l+1})=O(CC)$ and $v(C_{l+1}) + \sum_{C'
\in \Phi(CC \setminus C_{l+1})} P[C']$ must be evaluated in the
computation of $P[CC]$.
% Furthermore, each pair $C', C'' \in \mathbf{C}_{= l+1}$ must
%lie in different connected components of $CC(A \setminus \mathbf{C}_{\leq l})$,
% for otherwise, that would imply the existence of a path on the CSG graph between two nodes of level $l+1$ that does not pass through 
%a node of level $\leq l$. Such a path cannot exist, as those two nodes must lie
% on separate branches of the pseudotree.
Thus, $V_{l+1} \leq V_l$, and so, by the inductive hypothesis, $V_{l+1} \leq P[A]$.

%$CS$ should contain one
%coalition of level $l+1$, $C'_{l+1}$ , for each $C'\in CC(A \setminus
%\mathbf{C}_{\leq l})$. 
%Hence, for each $C'\in
%\mathbf{C}_{= l+1}$, the splitting $C'$ of $CC'$ is evaluated
%linking $\{\mathbf{C}_{\leq l}, C' ,
%CC(A \setminus \mathbf{C}_{\leq l}) \setminus C'\}$ to $CS_{\leq l}$. Doing
%it iteratively for each $C'\in \mathbf{C}_{= l+1}$ we end up building a path
%from $CS_{\leq l}$ to $CS_{\leq {l+1}} = \{\mathbf{C}_{\leq l+1}
%, CC(A \setminus \mathbf{C}_{\leq l+1})\}$.

\vspace{-0.1in}\section{Complexity Analysis}
\label{sec:complexity}

\noindent Next, we determine the complexity of DyPE 
and compare it to those of DyCE
and IDP. Notice that each of these algorithms, for each evaluated subproblem:
stores its value (and possibly its best space) and evaluates a number of
subspaces (with a linear number of operations per subspace).
Accordingly, we assess their complexity based
on the number of subproblems (memory requirements) and the number of subspaces
evaluated (computational requirements).
% Memory
%requirements of these algorithms are linear to the number of evaluated
% subproblems (for each subproblem each algorithm stores its value and its best subspace). On the other
%hand, the running time of these DP algorithms is determined by the
%number of evaluated subspaces (each subproblem is computed by evaluating a
% number of subspaces, doing a constant amount of work per subspace).
% Accordingly, we will
%assess the memory and computational requirements based on
%the number of evaluated subproblems and subspaces, respectively.
%In our analysis, we pay special attention to
%the classic and tree-restricted CSG cases in which DyPE is guaranteed a
%non-redundant search.

% \vspace{0.1in}\noindent Having analysed the complexity of Algorithm ,
%we can readily assess the complexity of DyPE as follows:
%analysing the particular cases in which DyPE is guaranteed to be efficient,
%namely in general and tree-restricted CSG problems.

\vspace{0.05in}\noindent \textbf{Memory.}
DyPE evaluates, in addition to the
grand coalition, one subproblem for $C\in F(G)$ such that $A
\setminus C$ is feasible and $C$ does not contain the root. This number is equal
to the number of feasible coalition structures composed of one (the grand coalition) or two coalitions ($\vert
\Pi^{A}_2\vert$). 
Thus, the memory requirements of DyPE are within $\mathcal{O}(\vert \Pi^{\vert
 A \vert}_2 \vert)$.
In classic CSG, $\vert
\Pi^{A}_2\vert$ is $2^{\vert A \vert-1}-1$.
%, so DyPE has memory
%requirements within $\mathcal{O}(2^{\vert A \vert})$. 
In tree-restricted CSG,
 $\vert \Pi^{A}_2\vert$ is equal to the number of edges in the tree 
(removing exactly one edge is the only way
to disconnect the tree into two connected subsets), so DyPE has memory
requirements within $\mathcal{O}(\vert A \vert)$.

Table \ref{tab:complexities} shows how the memory requirements of DyPE
compares to those of IDP and DyCE on graph-restricted CSG. Table
\ref{tab:complexities} also highlights the particular cases of classic and
tree-restricted CSG.
Observe that independently from the graph the complexity of IDP is exponential
in the number of agents, whereas of DyCE is linear in the number of feasible
coalitions. Since the number of feasible coalitions will always be greater than the number of coalition structures composed of two coalitions, the memory requirements of DyPE are bounded above by those
of DyCE. In classic CSG, although of the same order, the memory requirements
of DyPE are one half those of IDP or DyCE as among subproblems that contain the
root DyPE only stores the grand coalition.

\vspace{0.05in}\noindent \textbf{Computation.}
%DyPE evaluates subspaces of the grand coalition (that is one subspace for
%each feasible coalition that contains the root) and for each ($C_k$, $C_l$)$\in
%\Pi^A_2$  all subspaces of $C_k$, where $C_k$ is the part that does not contain
%the root (that is one subspace for each feasible coalition that contains the
%agent with lower order in $C_k$).
%In trees, one subspace for each feasible coalition that contains the root, are
%all that contain the root. Part that contains one agent and all down, are all
%coalitions that contain this agent as a higher level.
%For classic CSG, this involves
 DyPE evaluate subspaces which are subsets of subproblems. Thus, the
computational complexity is bounded by a constant times the number of pairs of
subsets $C', C$ with $C' \subseteq C$, which is $\mathcal{O}(3^{\vert A
\vert})$. This is the same order of complexity as IDP (and of DyCE in the
classic CSG). In classic CSG, DyPE omits the evaluation of all subsets of
subproblems that include the root node with exception of those of the grand
coalition. 
%that does not contain the root node are evaluated, along with the
%grand coalition as a subproblem, and all subsets of these subproblems are
% feasible.
Thus, we cannot hope to do better than $\mathcal{O}(3^{\vert A \vert})$.
In tree-restricted CSG, for each agent $i$, DyPE evaluates exactly one subproblem
with $i$ as its lowest order agent. Thus, each feasible coalition can only
generate a subspace in one subproblem, namely the one that has the same order.
Conversely, the subproblem with $i$ as its lowest order agent contains agent $i$
and all agents reachable from $i$ with higher level, and so all feasible
coalitions generate exactly one subspace.
% Conversely, the lowest level agent in a
%feasible coalition has to be an ancestor of all agents in that coalition, and
% so all feasible coalitions appear as splittings.
Thus, the computational
complexity of DyPE in this case is within $\mathcal{O}(\vert F(G)\vert)$, and
thus expected to be much lower than those of DyCE since the latter evaluates a
potentially large set of subspaces for each feasible coalition. 
%Comparing with IDP and DyCE, we observe the computational
%requirements of IDP are in $\mathcal{O}(3^{\vert A \vert})$ (independently of
%the graph). In classic CSG, DyCE also has this complexity. 
%For graph-restricted
%CSG, the complexity of DyCE depends on the underlying graph since it restricts
%to subspaces of evaluated subproblems that are produced by feasible partitions.
%Since the number of evaluated subproblems by DyPE is bounded above by those of
% DyCE, that same applies to its computational complexity.
%For tree-restricted CSG, DyPE evaluates a number of subspaces
%equal to the number of feasible coalitions whereas DyCE evaluates a potentially
%large set of subspaces for each feasible coalition. 
Indeed, in the next section
we show empirically that this holds true for a wide range of graph structures.

 \begin{table}[!tb]
\centering
\small
\begin{tabular}{|l| l| l| l|}
\hline
			& 	\textbf{Graph  (G)} & \textbf{Classic} &  \textbf{Tree  (T) }  \\
\hline
IDP & 
 $\mathcal{O}(2^{\vert A \vert} )$ & $\mathcal{O}(2^{\vert A \vert} )$ &
$\mathcal{O}(2^{\vert A \vert} )$  \\
\hline
DyCE &
$\mathcal{O}(\vert F(G) \vert)$ & $\mathcal{O}(2^{\vert A \vert})$ &
$\mathcal{O}(\vert F(T) \vert)$ \\
\hline
DyPE &  $\mathcal{O}(\vert \Pi^{\vert A
\vert}_2 \vert)$ & $\mathcal{O}(2^{\vert A \vert} )$   & $\mathcal{O}(\vert A
\vert)$ \\
\hline
\end{tabular}
\vspace{-0.05in}
\caption{Memory requirements for the graph-restricted, classic, and
tree-restricted CSG problems.\label{tab:complexities}}
\vspace{-0.05in}
\end{table}

%To assess the computational complexity of DyPE we need to
%assess the complexity of three processes, namely the complexity of: enumerating
%subproblems in $\mathbf{P}$, enumerating splittings in $\mathbf{S_C}$ and
%evaluating the value of an splitting.
%To enumerate the ordered set $\mathbf{P}$ and the set of splittings in
%$\mathbf{S_C}$, DyPE uses a CSE algorithm. The complexity of this CSE algorithm
%is linear to the number of feasible coalitions to be enumerated. 
% Thus, $O(E(\mathbf{P})) = O(\vert F(G) \vert) $ and $O(E(\mathbf{S})) =
 % O(\mathbf{S}) $.
%Finally, to compute the value of an splitting, DyPE requires the computation of
%the set of connected components in $C\setminus S$ which can be computed in
%linear time in terms of the number of agents and edges of the graph
%($\mathcal{O}(\vert A \vert \cdot \vert E \vert)$ ).
%Thus, the total
%computation time required by Fast-DP for the computation of the value of
%splittings is in $\mathcal{O}(\mathbf{S}\cdot \vert A \vert \cdot \vert E
%\vert)$.
%Notice that since the number of splittings will always be greater than the
%number of feasible coalitions, the computational complexity of $Fast-DP$ is
%linear to the number of evaluated splittings $\mathcal{O}(\vert
%\mathbf{S}\vert)$.
%Let's see which will be this number in particular graph cases.
%For trees, the number of splittings evaluated by Fast-DP is equal to the number
%of feasible coalitions $\mathcal{O}(\vert F(T) \vert)$. For complete graphs, 
% is $\frac{3^{\vert A
%\vert -1}-1}{2} + 2^{\vert A \vert
%-1} $.

\vspace{-0.1in}\section{Experimental Evaluation}
\label{sec:experimental}
\noindent We evaluate DyPE and compare its performance against
IDP and DyCE on a variety of different synergy graph topologies. We then go on
to examine the issue of scalability.

\begin{figure*}[!t]  
\begin{center}
	\subfloat[Memory requirements RT.
	\label{fig:results_random_memory}]{
		\includegraphics[width=0.45\textwidth,natwidth=551,natheight=364]{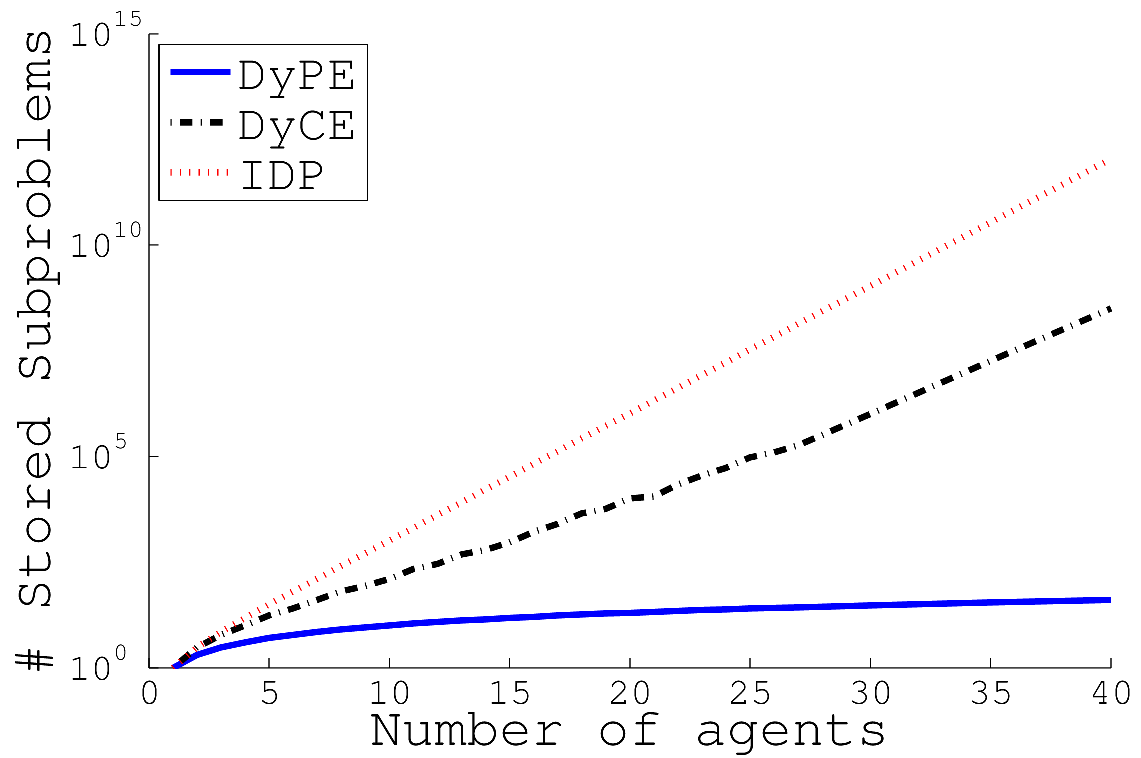}
		}
	\subfloat[Execution Time RT.
	\label{fig:results_random_execution}]{
		\includegraphics[width=0.45\textwidth,natwidth=548,natheight=364]{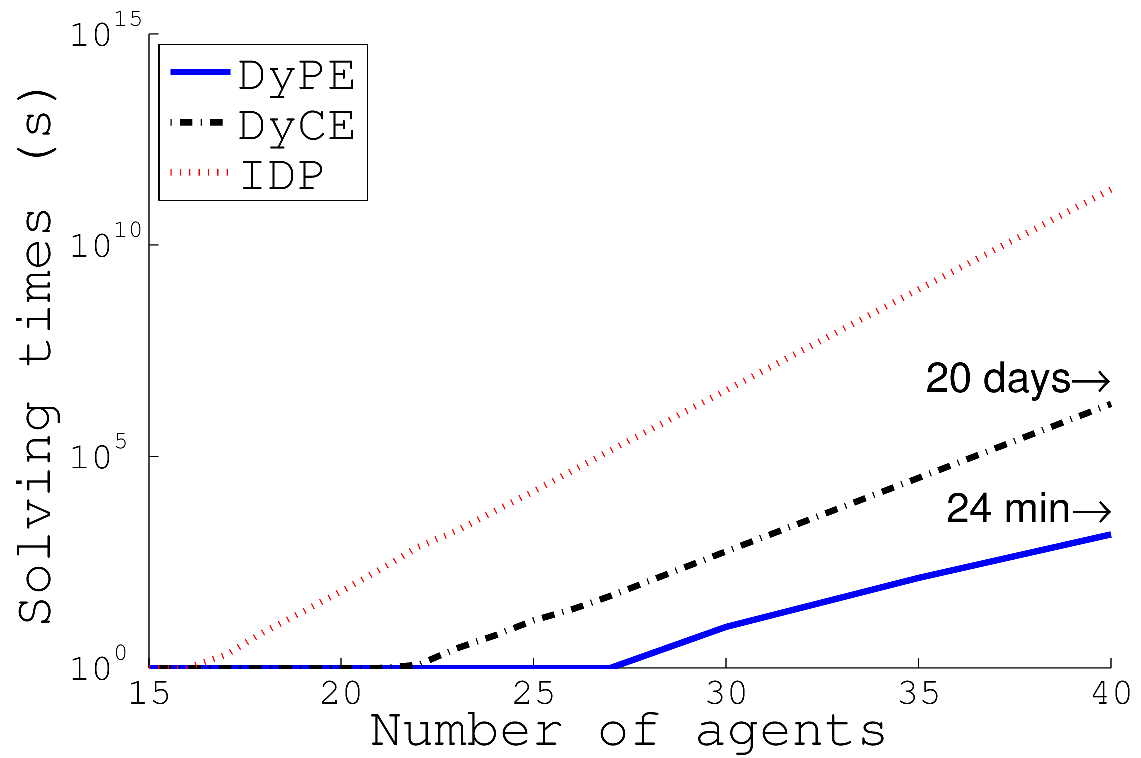}
		}

	\subfloat[Memory requirements SF.
	\label{fig:results_scalefree_memory}]{
		\includegraphics[width=0.45\textwidth,natwidth=548,natheight=364]{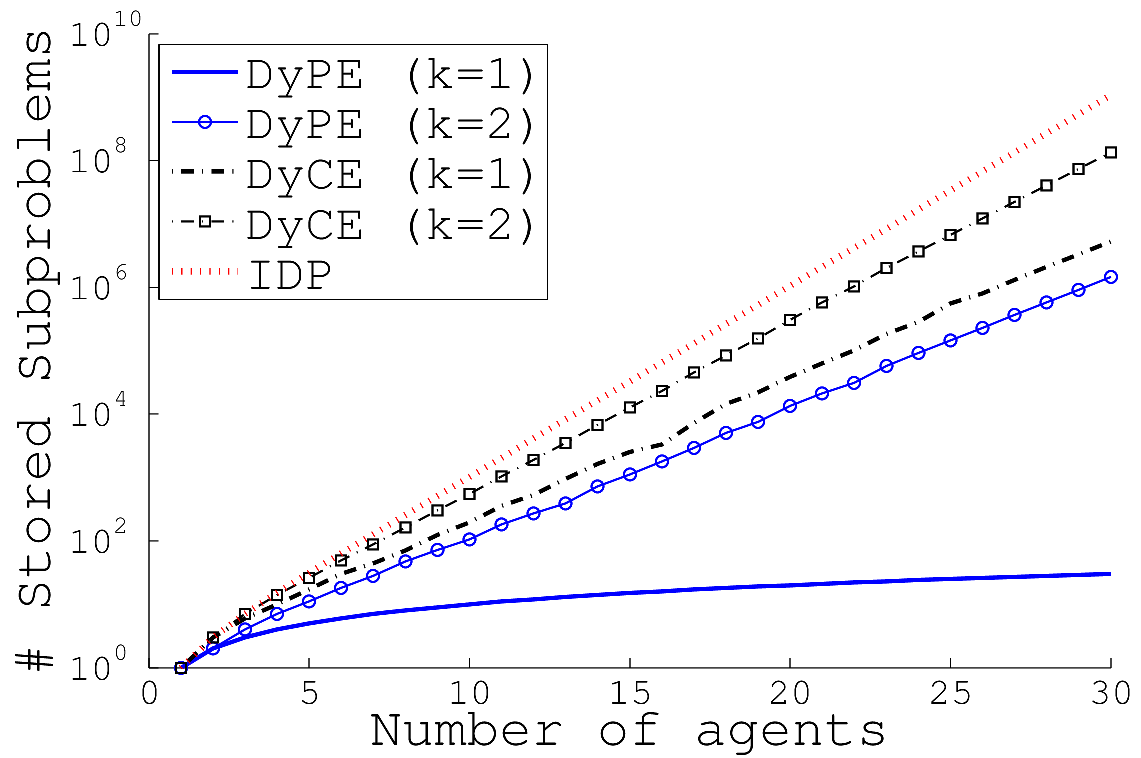}
		}
	\subfloat[Execution Time SF.
	\label{fig:results_scalefree_execution}]{
		\includegraphics[width=0.45\textwidth,natwidth=540,natheight=364]{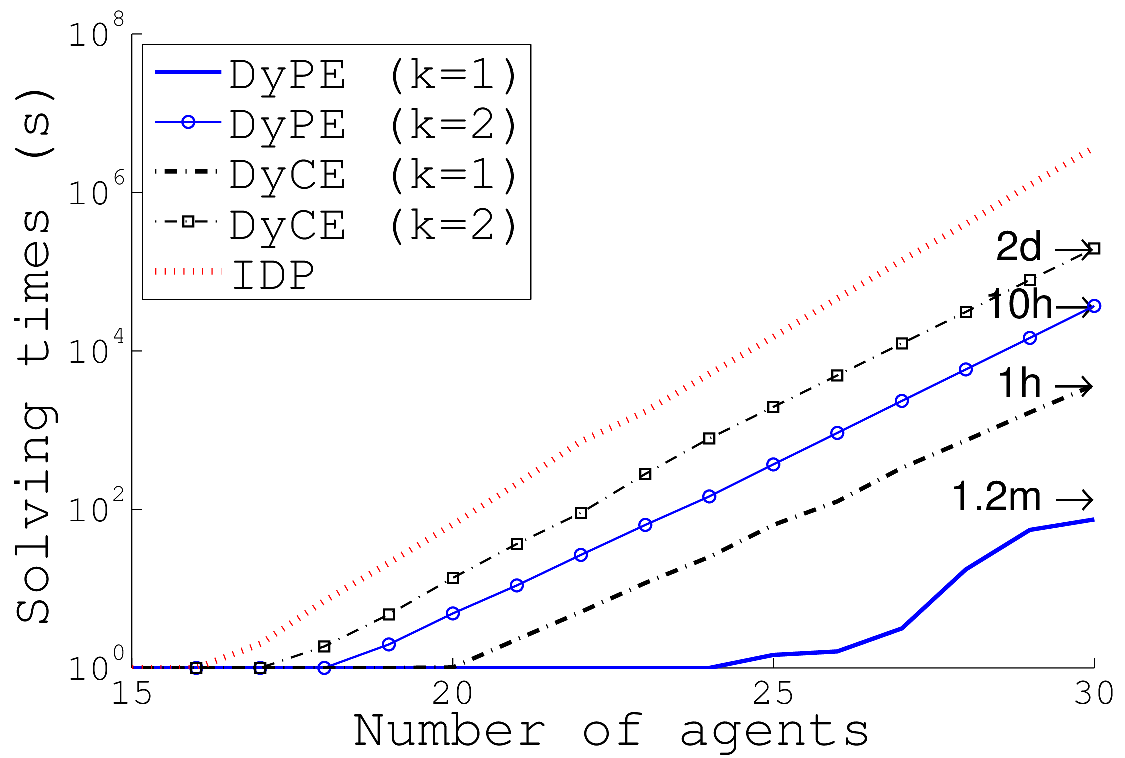}
		}
	\caption{Results for random tree (RT) graphs (a) (b) and scale free (SF) graphs
	(c)(d).}
	\label{fig:results}
\end{center}
\end{figure*}
\subsection{Benchmarking DyPE}
\label{sec:benchmarking}
\noindent 
In our comparison, we take a similar approach to \cite{DBLP:conf/aamas/VoiceRJ12}, 
and investigate performance over the following graph classes:
random trees (RT), scalefree graphs (SF) (using the standard Barabasi-Albert
preferential attachment generation model, with parameters $k=1,2,3$) and
complete graphs (CG).
Due to long runtimes we extrapolated the results as follows:  from 23 agents
onwards for IDP, from 27 onwards for DyCE on RT and SF $k=1$, and on 24 onwards
for DyCE on $k=2$. For each configuration, we run 50 instances
recording the number of evaluated subproblems and the running time of each
algorithm. We now present the results of this comparison.

 Figures \ref{fig:results} (a)-(b) show the results of our performance
evaluation over \emph{random trees}.
The memory requirements for DyPE are up to 7 orders of magnitude
lower than for DyCE and up to 11 orders of magnitude lower
than for IDP (for 40 agents).
This is because as the number of agents increases, memory requirements grow
exponentially for IDP and DyCE, and only linearly for DyPE. In terms of runtime,
DyPE can
solve problems of 40 agents in about 20 minutes compared against 20 days for
DyCE, and years for IDP. These results are in line with the intuition given by
our complexity analysis section.

The results of our performance evaluation over \emph{scalefree
graphs} are depicted in Figures \ref{fig:results} (c)-(d).
For  $k=1$ the memory requirements of DyPE for 30 agents are up to 6 orders of
magnitude lower than for DyCE and up to 9 orders lower than for IDP. For $k=2$,
these savings are reduced, but still significant; 2
orders of magnitude better with respect to DyCE and 3 orders of magnitude better with respect to IDP in graphs with 30 agents. 
These results follow the intuition given in our complexity analysis that the
computational savings provided by DyPE more significant on sparse graphs.
Turning to execution time, DyPE can solve problems with 30 agents in minutes (or hours when $k=2$) instead of hours (or days for $k=2$) for DyCE.

\noindent Finally, our results over \emph{complete graphs} are in line with the complexity analysis, which predicted a similar performance for all algorithms,
excepting that:
(i) DyCE takes more time than IDP and DyPE due to its less effective pruning;
and (ii) DyPE uses half of the memory of the other approaches.

\subsection{Scalability of DyPE}
\label{sec:benchmarkinglarge} 
    
\begin{figure}[!t]  
\begin{center}
	\includegraphics[scale=0.3,natwidth=551,natheight=364]{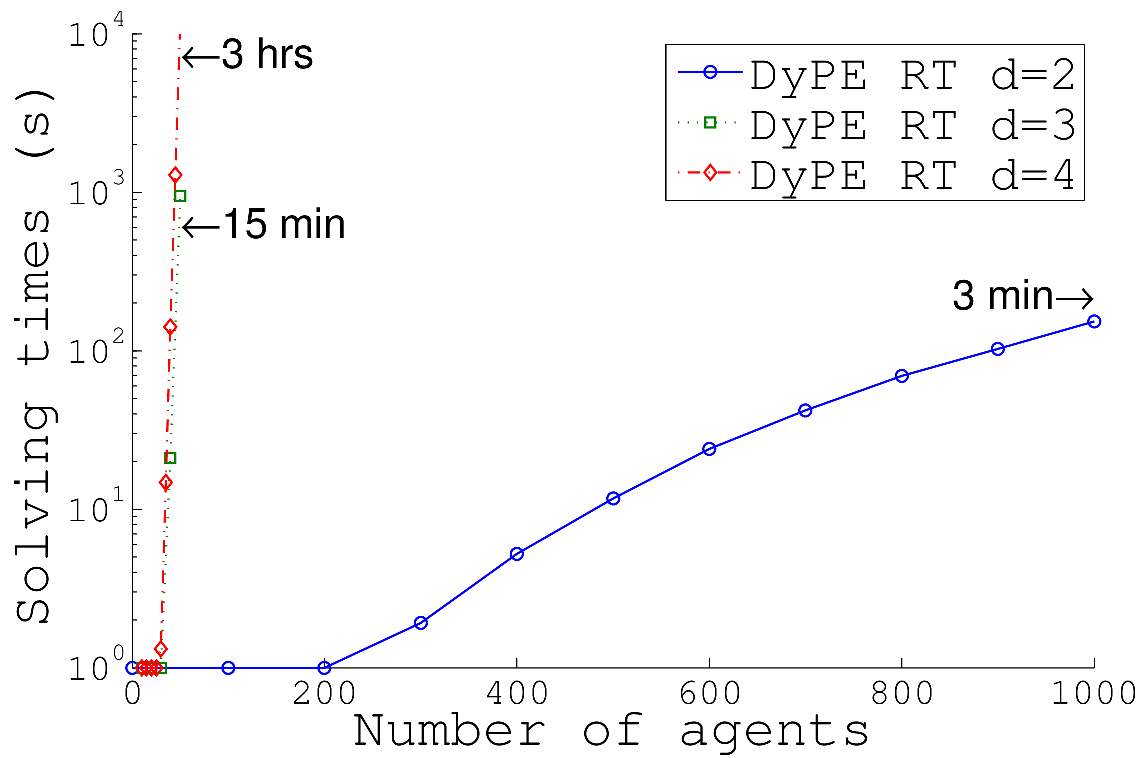}
	\caption{Runtimes for DyPE on random trees
	(d=2,3,4).\label{fig:results_largetrees}}   
\end{center}
\end{figure}

\noindent We have seen that DyPE performs well on sparse graphs (e.g., trees or
scale free with $k=1$). However, as argued in
\cite{DBLP:conf/aamas/VoiceRJ12}, if the degree
of agents is not bounded, even trees can lead to an
exponential number of coalitions (e.g., a star has $2^{\vert
A\vert-1}-1)$. Based on this, we evaluated DyPE on random trees with bounded
degree. In particular, Figure \ref{fig:results_largetrees} shows the execution
time where the degree of agents is bounded by $d$, for $d=2,3,4$. Observe that for $d=3$ and $d=4$, DyPE is able to run to completion
for problems with 50 agents within 15 minutes and 3 hours respectively. For the
particular case of $d=2$, DyPE solves problems with 1000 agents within
minutes.

\vspace{-0.05in}\section{Conclusions}
\label{sec:conclusions}
\noindent We presented DyPE, a DP algorithm that implements a
novel hierarchical DP formulation for CSG using a hierarchy based on
pseudotrees. We proved that DyPE is optimal and that it improves upon current DP
approaches with savings that go from linear to exponential, depending on the
structure of the underlying synergy graph.
%We proved that DyPE is optimal, and we showed that, unlike current DP
%approaches, for classic CSG or CSG restricted by a
%tree, DyPE is guaranteed a non-redundant exploration of the search space.
% any
%redundant computation or memory usage.
Our empirical results showed, that DyPE 
greatly improves on the state-of-the-art, in some cases by several orders of magnitude.
Concretely, for random trees with bounded
degree, DyPE managed to quickly find the optimal coalition structure for $1000$ agents,
when even $50$ would be intractable for other DP approaches.

As future work, following current trends in the field
\cite{DBLP:conf/aaai/RahwanMJ12,DBLP:journals/aamas/ServiceA11a},  we are
particularly interested in enhancing the presented hierarchical DP approach with
anytime properties.
%improve the average-case running times of the presented
%algorithms.
%allow returning solutions before completion with quality-guarantees and
% improved the average running time of.
% (either by combining it with anytime search algorithms as in or using
%the intermediate results of the DP algorithm to generate solutions anytime as
% in ) to improve the average-time

% Future work will look at combining DyPE with search algorithms (such as
% \cite{DBLP:conf/aaai/RahwanMJ12}), by extending the latter to search the space
 % of subspaces based on our hierarchical DP formulation.
%Regarding future work, given the recent success of hybrid approaches
%\cite{DBLP:conf/aaai/RahwanMJ12}, we are particularly interested in combining

%To date, the fastest approach to solve the CSG problem is an hybrid
%algorithm \cite{DBLP:conf/aaai/RahwanMJ1}, termed  IDP-IP*,  that merges
%together the advantages of dynamic programming and tree-search techniques by
% combining IDP and IP algorithms.
%However, as IDP, IP 
%divides the search space into subspaces that are searched by its size. 

\newpage
\bibliographystyle{aaai}
\bibliography{aaai2013} 
\end{document}